\documentstyle[preprint,aps]{revtex}

\begin{document}

\title{Single layer/bilayer transition of electron systems in
AlGaAs/GaAs/AlGaAs quantum wells subject to in-plane magnetic fields}

\draft
\author{L.\ Smr\v{c}ka, T.\ Jungwirth}

\address{{ \it Institute of Physics, Academy of Science of the Czech
Republic,} \\
{\it Cukrovarnick\'{a} 10, 162 00 Praha  6, Czech Republic}}

\date{J. Phys.: Condens. Matter, to be published}

\maketitle

\begin{abstract}
Equilibrium properties of electrons in double-heterojunction
AlGaAs\-/GaAs\-/AlGaAs structures are investigated theoretically,
using a full self-consistent numerical method. The transition
from single to bilayer electron systems is discussed for the case
of double-junction structures with varying distance between
interfaces. In strong in-plane magnetic fields we observed the
same type of transition accompanied by large effects on the
energy spectra. A new fenomenon in bilayer systems --- separability
of the energy dispersion curve in parts corresponding to
electrons either in first or second layer is analyzed in detail.
The magnetic field induced variation in the effective distance
between electron layers is compared to the constant layers
separation in standard double-well structures. Due to relatively
small and soft electrostatic barrier between interfaces we expect
the van Hove singularity in the 2D density of states to be easier
detectable than in double-well systems. The cyclotron mass,
proportional to the density of states and characterizing the
electron motion in tilted magnetic fields, is calculated as
a function of in-plane magnetic field. The separability of the
energy spectra resulting in tilted cyclotron orbits is
demonstrated on a classical picture of an electron moving in
crossed electric and magnetic fields.
\end{abstract}


\narrowtext

\section{Introduction}
The electronic structure of bilayer 2D electron systems in double
quantum wells is in the focus of interest since their preparation
and intensive theoretical and experimental investigation of their
properties is now in progress. Main directions of recent research
of these structures subjected to magnetic fields were reviewed by
Eisenstein \cite{1}. Firstly, in a strong perpendicular magnetic
field, the double layer systems exhibit new quantum Hall states
not present in a standard 2D gas because of an extra degree of
freedom, the layer index. Secondly, in the in-plane magnetic
fields, the tunneling between two electron layers can be modified
to such extent  that, at certain critical fields, the interlayer
tunneling is completely suppressed and electrons move exclusively
in one of the wells. The field-induced decoupling of electron
layers is accompanied by a number of new phenomena which can be
studied by measuring both the parallel and perpendicular
transport and the Shubnikov-de Haas effect.

A typical double well structure consists of a pair of narrow
wells ($\approx 20$ nm) separated by an undoped AlGaAs barrier of
a similar thickness. Two Si doping layers are usually located
both below and above the wells and the Schottky gates are
deposited on the top/back of the sample to control better the
concentration of electrons in individual wells. Due to the
tunneling through the barrier the degeneration of the energy
spectrum in the layer index is removed, the corresponding single
levels split and the lowest bound states of individual wells form
a symmetric/antisymmetric pair. A small tunneling probability
through the hard-wall AlGaAs barrier implies the small level
splitting and the weak coupling of electron layers.

Traditionally, the gross features of the electronic structure of
a standard double well are explained in terms of a simple model
which assumes two identical very narrow wells at the distance
$d$ and a weak overlap of the localized states corresponding to
first subbands of both wells. The interwell tunneling is
described in the tight binding approximation by a single
parameter determining a small bonding-antibonding energy gap.
Then the band spectrum is given by the elementary nearly free
electron model as presented in classical solid state textbooks
(see e.g. \cite{Mott}). The model predicts the anticrossing of
two circular Fermi lines accompanied by the logarithmic van Hove
singularity in the density of states which can be interpreted, in
the 2D case, in terms of singular effective cyclotron mass. It
was utilized to study a role of minigaps in Si by Ando
\cite{Ando} and Mathesson and Higgins \cite{Math} and
successfully applied for description of the interwell tunneling
(see \cite{1} and references therein) and of the electronic
structure of double wells in tilted magnetic fields \cite{Hu},
\cite{Lyo} and \cite{tj}.

In our theoretical study we shall consider another potentially
bilayer systems, the double-heterojunction quantum wells formed
only by two selectively doped AlGaAs/GaAs interfaces with no
AlGaAs layer grown between them. Note that in this case an
undoped structure would result in an exactly rectangular quantum
well. In the doped structures, the transfer of electrons from Si
donors into the GaAs leads to the band bending in the GaAs by
electrostatic forces and, consequently, to the formation of
a soft build-in barrier of an approximately parabolic shape
inside the well. It depends on the well width and the amount of
the charge wether the built-in barrier is strong enough to
separate the quantum well into two independent/weakly-coupled
parts or if the coupling is so strong that the system can be
considered as a transient form between a single well and a double
well. For simplicity, only the symmetrically doped wells will be
considered.

In charged wells, the shape of the barrier is given by the charge
distribution. Therefore the electronic structure must be
determined by the self-consistent solution of coupled
Schr\"{o}dinger and Poisson equations. There is a number of such
calculations for the case of zero magnetic field, see e.g.
\cite{2} and references therein. Effects of the in-plane magnetic
field on both conduction and valence bands, resulting in
a diamagnetic shift of the electron-hole recombination energy,
were discussed by Oliviera {\it et al.} \cite{3}. Here the full
self-consistent calculation will be employed to obtain the
electron energy spectra and information about the charge
redistribution due to the magnetic field. The electronic
structure of double-heterojunction wells will be studied mainly
from the point of view of the possible magnetic field induced
splitting of a single layer electron system to a bilayer one.

\section{Band structure calculation}
The standard semi-empirical model working quantitatively for the
lowest conduction states of GaAs/AlGaAs heterostructures is used
to solve the Schr\"{o}dinger equation in the envelope function
approximation. The envelope function is assumed to be built from
host quantum states belonging to a single parabolic band. The
effect of the effective mass mismatch is completely neglected and
the envelope functions of GaAs and AlGaAs are smoothly matched at
the interface. Due to the translational invariance in the layer
plane the momentum operators $p_x$ and $p_y$ become good quantum
numbers, $p_x \rightarrow \hbar k_x$ and $p_y \rightarrow \hbar
k_y$. If we choose a vector potential ${\bf A}$ in the form
$ {\bf A} = (B_yz,0,0)$, the eigenvalue problem reduces to solving
the one-dimensional Schr\"{o}dinger equation with the Hamiltonian

\begin{equation}
\label{a}
H =  \frac{1}{2m}p_z^2 + \frac{m\omega^2}{2} (z -z_0)^2 + V_{conf}(z).
\end{equation}

\noindent For more detailed derivation, consistent with our
notation, see e.g. \cite{Smr} and references therein.
The centre $z_0$ of the the magnetic part of the
effective potential energy is related to the wave vector
component $k_x$ by $z_0 = \hbar k_x/m\omega$, $\omega$ denotes
the cyclotron frequency $\omega = |e|B_y/m$. The confining
potential energy of a well with the width $d$

\begin{equation}
V_{conf}(z)=V_b(z)+V_{sc}(z)
\end{equation}

\noindent is a sum of the step functions
$V_b(z)=V_b\,\theta(-z-d/2) + V_b\,\theta(z-d/2)$ corresponding
to the conduction band discontinuities between AlGaAs and GaAs
and of a term $V_{sc}(z)$ describing the interaction of an
electron with ions and the electron-electron interaction. This
term should be calculated self-consistently and can be written as

\begin{equation}
V_{sc}(z)=V_H(z)+V_{xc}(z).
\end{equation}

\noindent The Hartree term $V_H$ is the solution to the Poisson
equation

\begin{equation}
\label{b}
\frac{d^{2}V_{H}}{dz^{2}}=\frac{|e|\varrho(z)}{\epsilon}
\end{equation}

\noindent and we use an expression calculated by Ruden and
D\"{o}hler \cite{Ruden} in a density-functional formalism for the
exchange-correlation term $V_{xc}$,

\begin{equation}
V_{xc}\simeq-0.611\frac{e^2}{4\pi\epsilon}
\left(\frac{3N_e(z)}{4\pi}\right)^{1/3}.
\end{equation}

\noindent The conduction band offset $V_b$ and the dielectric
constant $\epsilon$ enter our calculations as input parameters.

For modulation doped GaAs/AlGaAs heterostructures the total
charge density $\varrho(z)$ in the equation (\ref{b}) can be
split into parts corresponding to concentrations of electrons,
$N_e(z)$, their parent donors in AlGaAs, $N_d^+(z)$, and ionized
residual acceptors in GaAs, $N_a^-(z)$,

\begin{equation}
\varrho(z)=e\left[N_e(z)-N_d^+(z)+N_a^-(z)\right].
\end{equation}

\noindent We accept a usual approximation of constant impurity
concentrations and assume donors and acceptors to be ionized
within certain finite intervals: $N_d^+(z)=N_d$ for $d/2+w\leq
|z|\leq d/2+w+l_d$, i.e. symmetrically with respect to the well
of the width $d$, and $N_a^-(z)=N_a$ for $|z| \leq d/2$, which
means that all acceptors inside the well are ionized. Here $w$
denotes the spacer thickness and $l_d$ is the depletion length of
donors which is determined in the course of the self-consistent
procedure.

\section{Results and discussion}
In our calculation we consider four wells with widths $d$ = 25,
40, 50 and 60 nm, close to the dimensions of a typical double
quantum well described above. The concentrations of localized
charges are $N_d=2\times 10^{18}$ $\rm cm^{-3}$, $N_a=10^{14}$
$\rm cm^{-3}$, the band offset $V_b=225$ $\rm meV$ and the
dielectric constant $\epsilon = 12.9$. The spacer thickness $w$
= 60 nm yields electron systems with the concentration of
electrons $N_{e}\approx 3.2\times 10^{11}$ $\rm cm^{-2}$. The
self-consistent procedure was performed in two steps. Firstly,
the electronic structure of wells was calculated in zero magnetic
field, in the second step the magnetic field dependence of
subbands was obtained starting from intermediate zero-field
results.

\subsection{Zero magnetic field case}
It follows from the form of Hamiltonian (\ref{a}) that, in the
case of zero magnetic field, an electron motion in the $z$
direction is completely decoupled from the in-plane components.
The energy spectrum consists of two quasi-continuous free
electron branches, the quadratic functions of the wave vector
components $k_x, k_y$, and the energy levels corresponding to the
bound states of a well. Positions of levels on energy scale
are presented in figure \ref{f1}, together with the band diagrams
resulting from the first step of the self-consistent procedure.
The form of the barriers is approximately parabolic for all four
wells, the minor deviations from this shape may be observed for
 wider wells. The heights of barriers are very small when
compared with hard walls of  wells formed by the GaAs/AlGaAs
interfaces. This indicates a stronger coupling between wells than
in a standard double-well structure.

All the systems have one occupied pair of
symmetric/antisymmetric subbands, except of the 25 nm wide
well, where only the subband belonging to the lowest (bonding)
state is occupied. Note that the origin of the energy scale in
figure \ref{f1} is located at the Fermi energy $E_F$ of the
electron system. The energy spectrum of the 60 nm system
corresponds to almost independent quantum wells, both occupied
levels lie below the top of the barrier and the splitting of
energy levels due to the quantum tunneling is small. The coupling
of 2D electron layers increases smoothly with the decreasing well
width and the energy spectrum of the 25 nm well reminds rather of
a single well than a double well since the energy difference
between the bonding and the antibonding states is comparable with
the difference between the energy of the first level and a bottom
of the well.

The electron charge density $N_e(z)$ corresponds to well known
behaviour of two coupled wells as described by simple
tight-binding model: there are two maxima of the density
positioned symmetrically in each well, with the small but finite
minimum at $z=0$ for bonding states and $N_e(0)=0$ for
antibonding states. Of course, the maxima are closer and bonding
state charge densities higher in narrower double wells.

\subsection{Magnetic field dependent band structure}
As seen in figure \ref{f1}, it is possible to approximate the
form of barriers by an expression $-m\Omega^2 z^2/2$, where
$\Omega $ is the fitting parameter. Assuming further that the
barriers are not substantially changed by redistribution of
electrons in the magnetic field, the potential energy in
(\ref{a}) is given by a combination of the fixed electric barrier
and the attractive magnetic well with the curvature given by
$B_y$. The position of minimum, $z_0$, which need not lie inside
the interval $-d/2,+d/2$, is determined by $k_x$. For small
magnetic fields the potential energy inside the well is still
a barrier with a smaller curvature $m(\omega^2 - \Omega^2)$ and
a maximum at $z = z_0\,\omega^2/(\omega^2 - \Omega^2)$. At
a critical magnetic field, $B_y=m\Omega/|e|$, the bottom of the
well becomes a linear function of $z$, with the slope dependent
on $k_x$, and at higher fields the sign of curvature of the
potential energy line is reversed and a minimum appears at $z$
instead of a maximum. The critical fields for the considered
wells $B_{60}=4.7$ T, $B_{50}=5.6$ T, $B_{40}=6.7$ T and
$B_{25}=9$ T are fully in the range of experimentally available
values. This distinguishes the double-heterojunction structures
from the standard double wells with steep AlGaAs barriers in
which two deep minima are preserved for arbitrary magnetic
fields. Consequently, the simple tight binding picture of
eigenfunctions as symmetric/antisymmetric combinations of the
eigenstates of individual wells is broken. Above a critical field
the lowest eigenstates exhibit always a single maximum and thus
electrons are more localized by the magnetic field around its
centroid then in the case of standard structures. Examples of
eigenfunctions together with \lq electro-magnetic\rq{} potential
energy lines defined by equation (\ref{a}) are presented in
figure \ref{extra}. The results are plotted for selected values
of $B_y$ and, to demonstrate their $k_x$ dependence, also for
a minimum and a maximum $|k_x|$ corresponding to occupied states.

The transition from a single layer to bilayer electron system,
induced by the in-plane magnetic field, is clearly visible in
figure \ref{extra}. Moreover, the effective inter-layer
separation increases with the magnetic field reaching a value
equivalent to the width of the central GaAs region, in the strong
magnetic field limit. These effects are absent in standard double
well systems where the distance between 2D electron layers is
a constant given by the separation between wells.

It follows from the above discussion that the in-plane magnetic
field does not influence the harmonic dependence of subbands on
$k_y$ at all, but the dependence on $k_x$ is changed
dramatically. Firstly, the energy separation of bottoms of the
bonding/antibonding pair of subbands increases with growing
$B_y$. This process is accompanied by the continuous transfer of
electrons from the higher occupied subband to the lower one until
the second subband is completely depleted. Secondly, the
curvature of the originally parabolic subbands is modified. The
antibonding subband remains almost parabolic, similarly as in
zero field, only its curvature is slightly higher for larger
$B_y$. The bonding subband exhibits more varied field dependence.
The curvature of $E(k_x)$ decreases for $k_x$ around $k_x = 0$,
at certain value of $k_x$ reaches zero and then becomes negative.
This means that, instead of a minimum which corresponds to a zero
field, the local maximum develops at $k_x = 0$, accompanied by
two new minima positioned symmetrically around it. For large
enough magnetic fields the energy of the maximum becomes greater
than the Fermi energy and the spectrum of occupied energies
splits into two separated parts. This type of behaviour is
general for energy spectra of all four well widths. In the narrowest
well the splitting does not occur in the investigated range of
fields. Examples of $E(k_x)$ curves for the well of a medium
width, $d$ = 40 nm, and several selected values of $B_y$ are
presented in figure \ref{f2}.

The electrons with energies close to the Fermi energy $E_F$ are
of particular interest since they play essential role in the
electronic transport properties of the system. The occupied and
empty states are separated by the Fermi lines in $\bf k$ space.
In a zero magnetic field, the Fermi contours corresponding to
symmetric and antisymmetric subbands have a form of two
concentric circles. The larger circle  describes the bonding
and the smaller circle the antibonding states. The modification of
the shape of $E(k_x)$ curves by the in-plane magnetic fields is
reflected also in variation of forms of Fermi contours. They
acquire a \lq peanut\rq{} and a \lq lens\rq{} shapes when the
sample is subjected to the intermediate magnetic fields. In the
strong field limit the \lq lens\rq{} is emptied and the \lq
peanut\rq{} is split into two parts belonging to individual
wells. This behaviour is illustrated in figure \ref{f3} for the
same electron system and the same set of magnetic field values as
in figure \ref{f2}. The deviations from two crossing circles,
corresponding to a pair of narrow independent wells,
are so large that the shape of Fermi lines can hardly be
described by the above mentioned simple model.

\subsection{Cyclotron effective mass and real space
trajectories}
The separation of electron layers by the in-plane magnetic field
may be clearly demonstrated if we consider the electronic
structure of a system subjected to a tilted magnetic field with
a strong component $B_y$, parallel to the GaAs/AlGaAs interfaces,
and a weak component $B_z$ oriented perpendicularly. To describe
the system, the self-consistent quantum mechanical calculation of
the electron subbands in the presence of $B_y$, presented above,
is combined with a quasi-classical description of the in plane
electron motion under the influence of $B_z$, as developed
originally by Onsager \cite{Ons} and Lifshitz \cite{Lif}.

The semiclassical theory predicts that an electron is driven by
the Lorentz force due to $B_z$ around the Fermi contour with the
cyclotron frequency $\omega_c = |e|B_z/m_c$. The cyclotron
effective mass $m_c$, defined by the above equation, is an
important characteristics of each Fermi line as a whole and
should not be confused with the electron effective mass which is
in our case a tensor and an anisotropic function of the position
on the Fermi line. The explicit expression relating $m_c$ to the
shape of the Fermi contour is

\begin{equation}
\label{c}
m_c = \frac{\hbar^2}{2\pi}\, \oint \frac{dk}{|\nabla_kE|}\,,
\end{equation}

\noindent where $dk$ denotes an element of a length of a Fermi
line. In the case of two-dimensional electron systems a simple
relation holds between the cyclotron effective mass and the
density of states $g$ corresponding to a single subband:

\begin{equation}
\label{d}
g = \frac{m_c}{\pi \hbar^2}\, .
\end{equation}

\noindent Note that this is the cyclotron effective mass, and not
the effective mass, which determines e.g. the temperature damping
of Shubnikov-de Haas oscillations.

Both equations (\ref{c}) and (\ref{d}) can be found, at least in
an implicit form, in \cite{Ons} and \cite{Lif} already. Since
then, they were utilized many times, for application to 2D
systems see e.g. \cite{Math}, but overlooked in the recent
publication by Harff {\it et al.} \cite{Harff}. Their detailed
derivation and discussion for the case of combined parallel and
perpendicular magnetic fields is presented in \cite{Smr}.

The magnetic-field dependent cyclotron effective masses resulting
from our band structure calculation are shown in figure \ref{f4}
for all four wells. The logarithmic singularities of the
cyclotron mass (density of states) curves corresponding to the
lowest subbands occur at higher fields for narrower wells than
for broader ones. The position of a singularity is identical with
the field value at which a \lq peanut\rq{} becomes a pair of \lq
kissing\rq{} Fermi contours, i.e. when it is to split into two
independent parts. Note that this is always above the
fields $B_{60} = 4.7$ T, $B_{50} = 5.6$ T and $B_{40} = 6.7$
T discussed above. The asymmetric shape of van Hove singularities
somewhat deviates from $-ln|E_F(B_y)-E_0|$ predicted by the
analysis based on the quadratic approximation of the energy
spectrum near the saddle point. The reason for the slow increase
of $m_c$ on the left side from the singularity, and the rapid
decrease on the right side , may be due to the large influence of
higher order terms in the power expansion of $E(k_x)$. For the
Fermi contours of the second subband the cyclotron mass always
decreases for the given concentration of electrons.

In the real space, electrons move along trajectories having the
shape of Fermi contours but rotated by an angle $\pi/2$ and
multiplied by $\hbar/|e|B_z$. A relation exists between the
centre of mass $\langle z \rangle$ of an electron state in a well
and the energy spectrum $E(k_x)$ modified by the in-plane
magnetic field,

\begin{equation}
\langle z \rangle = \frac{\hbar k_x}{m\omega} - \frac{1}{\hbar
\omega}\, \frac{\partial E(k_x)}{\partial k_x}\,.
\end{equation}

\noindent Since $\langle z \rangle$ is a function of $k_x$, i.e.
of the position of an electron on the Fermi contour, the real
space trajectory is slightly tilted with respect to the $x-y$
plane. Travelling around a \lq peanut\rq{} Fermi contour with the
frequency $\omega_c$, the electron is transmitted from one well
to the other when the wavevector $k_x$ changes its sign. At $k_x
= 0$ the electron appears on the top of the barrier. The
separation of a \lq peanut\rq{} into two parts means that two
independent real space trajectories for electrons are formed,
localized completely in opposite wells. In that case, the
electron moving around a Fermi contour is always reflected by the
barrier and remains confined in one of the wells. Hence, the
electron system can be considered as divided into two
disconnected layers. Self-consistently calculated examples of
$\langle z \rangle$ as a function of $k_x$ are presented in
figure \ref{f5}.

The quasi-classical model of an electron motion in a double well
can help to illustrate the origin of localization of electrons in
individual wells due to the magnetic field. Figure \ref{f6} shows
real space trajectories of an electron in a well with a parabolic
barrier calculated classically for the case of two tilted
magnetic fields with the same perpendicular component. For the
field below the singularity, the trajectory overcomes the barrier
and an electron spends the same time in both wells. Note that, in
accord with this statement, the left part of the curve (a) lies
above and the right part below the projection of the trajectory
to the $x-y$ plane, ($ z=0$). The curve (b) is the electron
trajectory calculated for the field above the singularity (the
curves (a) and (b) are offset for clarity). In that case, an
electron is kept all the time in one well. Just at the magnetic
field, when the singularity occurs, electron stops at the top of
the barrier and stays there infinitely long. It means, in terms
utilized above, that its cyclotron frequency is zero and the
cyclotron mass goes to infinity.

\section{Conclusion}
We have performed self-consistent calculations of the electronic
structure of symmetrically charged double-heterojunction wells in
parallel magnetic fields. Their properties are to the large
extend similar to standard double wells but, on the other hand,
there are several differences between them which make the
double-heterojunctions attractive.

In both types of structures an in-plane magnetic field
substantially modifies the $k_x$ dependence of energy subbands.
The following properties result from these field induced changes.
(i) Electrons are transferred from the antibonding to the bonding
band when magnetic field increases. (ii) The Fermi contour
topology changes, the originally circular lines acquire a \lq
peanut\rq{} and a \lq lens\rq{} shapes and, at certain critical
fields, the \lq lens\rq{} is emptied ({\em first critical field}) and
the \lq peanut\rq{} splits into two independent lines ({\em second
critical field}). (iii) The splitting of the \lq peanut\rq{} Fermi
contour is accompanied by the separation of an electron layer
into two independent systems. (iv) At the second critical field
the density of electron states and the cyclotron effective mass
exhibit a logarithmic singularity when considered as functions of
the in-plane field.

The hard wall barrier formed by the AlGaAs layer means that the
coupling between electrons in a standard double well is always
weak. Therefore, a simple tight binding method for description of
an electron motion in the $z$ direction, resulting in a nearly
free electron model for its motion in the $x-y$ plane, can be
successfully applied to describe the electronic structure.

This is not true for double-heterojunction systems. Due to the
charge present inside wells, a soft barrier of approximately
parabolic shape is built in each well, with the top in its
centre. The height of a barrier is very small and the strength of
coupling can be tuned in large range by proper choice of the
concentration of carriers and the width of the well. Under these
circumstances, the simple model fails to describe the electronic
spectra even in a qualitative way. Note, that according to our
calculations, e.g. the gap between two occupied subbans increases
under the influence of the field, while the simple model yields
a constant gap.

The strong coupling between wells means that stronger magnetic
fields are necessary to modify the electronic structure of
a double-heterojunction. The difference between the first and
second critical fields can amount several Tesla and,
consequently, also the changes of the cyclotron effective mass
are more robust and probably easier to detect experimentally.

Nevertheless, the above described differences between two types
of structures are mainly quantitative. The qualitatively new
features of double-heterojunction structures are the single
layer/bilayer transition and the possible control of the inter-layer
separation by in-plane magnetic field. These effects might be
useful in studies of many-body effects in bilayer electron
systems.

For these properties the double-heterojunction structures
represent a very attractive alternative to standard double wells.

\acknowledgments
This work has  been supported by the Academy of Science of the
Czech Republic under Grant No. 110414, by the Grant Agency of the
Czech Republic under Grant No. 202/94/1278, by the Ministry of
Education, Czech Republic under contract No. V091 and by NSF,
U.\ S., through the grant NSF INT-9106888.

\begin{figure}
\caption{Band diagrams and energy levels of double-heterojunction
wells at a zero magnetic field.}
\label{f1}
\end{figure}

\begin{figure}
\caption{Wave functions and \lq electro-magnetic\rq{} band
diagrams of a double-heterojunction well subject to in-plane
magnetic fields. Pairs of curves characterize the electrons
with a minimum (dashed)  and a maximum (full) occupied $|k_x|$.
The width of the well is 40 nm.}
\label{extra}
\end{figure}

\begin{figure}
\caption{Occupied energy subbands of a double-heterojunction
well, $d =$ 40 nm,
subject to in-plane magnetic fields as functions of $k_x$.}
\label{f2}
\end{figure}

\begin{figure}
\caption{Fermi contours of a double-heterojunction well,
$d =$ 40 nm, subject to in-plane magnetic fields.}
\label{f3}
\end{figure}

\begin{figure}
\caption{Cyclotron effective masses of four wells as functions of
an in-plane  magnetic field.}
\label{f4}
\end{figure}

\begin{figure}
\caption{Centroids of electron states of a double-heterojunction
well, $d =$ 40 nm, subject to in-plane magnetic fields.}
\label{f5}
\end{figure}

\begin{figure}
\caption{Classically calculated trajectories of an electron in
a double well subject to tilted magnetic field. The upper curve
(a) corresponds to $B_y$ below the singularity and the lower
curve (b) to $B_y$ above this value.}
\label{f6}
\end{figure}

\end{document}